\documentclass{article}

\usepackage{microtype}
\usepackage{amsmath}
\usepackage{graphicx}
\usepackage{subfigure}
\usepackage{booktabs} 

\usepackage{hyperref}

\usepackage[accepted]{icml2021}

\icmltitlerunning{Neural message passing for joint paratope-epitope prediction}

\begin{document}

\twocolumn[
\icmltitle{Neural message passing for joint paratope-epitope prediction}



\icmlsetsymbol{equal}{*}

\begin{icmlauthorlist}
\icmlauthor{Alice Del Vecchio}{cam}
\icmlauthor{Andreea Deac}{mila,udem,dm}
\icmlauthor{Pietro Li\`{o}}{cam}
\icmlauthor{Petar Veli\v{c}kovi\'{c}}{dm}
\end{icmlauthorlist}

\icmlaffiliation{cam}{University of Cambridge}
\icmlaffiliation{mila}{Mila}
\icmlaffiliation{udem}{Universit\'{e} de Montr\'{e}al}
\icmlaffiliation{dm}{DeepMind}

\icmlcorrespondingauthor{Alice Del Vecchio}{ard69@cam.ac.uk}

\icmlkeywords{Machine Learning, ICML}

\vskip 0.3in
]



\printAffiliationsAndNotice{} 

\begin{abstract}
Antibodies are proteins in the immune system which bind to antigens to detect and neutralise them. The binding sites in an antibody-antigen interaction are known as the paratope and epitope, respectively, and the prediction of these regions is key to vaccine and synthetic antibody development. Contrary to prior art, we argue that paratope and epitope predictors require asymmetric treatment, and propose distinct neural message passing architectures that are geared towards the specific aspects of paratope and epitope prediction, respectively. We obtain significant improvements on both tasks, setting the new state-of-the-art and recovering favourable qualitative predictions on antigens of relevance to COVID-19.
\end{abstract}
\section{Introduction}
\label{submission}

Antibody binding site prediction is an important unsolved problem in immunology and is a prerequisite for in silico vaccine and synthetic antibody design. Currently, the most accurate method for determining these regions is by observing the residues which are spatially close in the 3D bound structure, obtainable by experimental techniques such as X-ray crystallography \cite{smyth2000x}. However, these experimental methods are time consuming and expensive, so there is a need to develop computational methods which can overcome these problems to aid faster development of therapeutics. For example, in the ongoing COVID-19 pandemic, mutations in the antigen have been shown to cause changes to the binding mechanism \citep{covidmutant}, potentially impacting the efficacy of existing therapeutics. Such a computational tool would allow for quick determination of the impact of these mutations, allowing for therapeutics development to keep up with fast occurring mutants.

We only consider scenarios where the antigen is a protein. Hence, both the antibody and the antigen may be viewed as \emph{sequences} of amino acid residues. Their binding site consists of two regions: the \emph{paratope} on the antibody, and the \emph{epitope} on its corresponding antigen. Predicting them can therefore be posed as a binary classification problem: for each amino acid residue in the antibody and antigen, respectively, do they participate in the binding? 

However, proteins can also be considered as \emph{graphs} with its residues as nodes, with two nodes sharing an edge if their residues are spatially close. Recently, such contact graphs have been directly leveraged for protein function prediction by \citet{gligorijevic2020structure}.

The advantage of considering a sequence based approach over a graph-based approach is that structural information is much harder to obtain. However, recent advancements in single-protein structure prediction from AlphaFold 2 \citep{alphafold2} have comparable accuracy to experimental methods. This makes the incorporation of structure justifiable as obtaining this information is becoming easier.

In this work we study the joint epitope-paratope prediction task in detail, identify an inherent asymmetry between the two tasks, and propose \emph{epitope-paratope message passing} (EPMP), a hybrid method that explicity leverages this asymmetry to design potent predictors for both tasks. We set the new state-of-the-art on both tasks, and illustrate qualitatively meaningful predictions on antigens of present importance.

\section{Related work}

Of the two tasks, paratope prediction has seen much more success. This is because the paratope is necessarily tightly localised on the tips of the Y-shaped antibody (the so-called \emph{\textbf{Fv} region}). In fact, many early works on paratope prediction (e.g. \citet{parapred}) consider an even narrower region, the complementarity-determining regions (CDRs)---which combined rarely exceeds 180 residues. This allowed for discarding most of the antibody sequence, leading to a better-balanced prediction problem. In rare cases, the paratope can be located outside of the CDRs; accordingly, we will utilise the entire Fv region as input.

Paratope prediction was tackled both by sequential \citep{parapred,agfastparapred,multitask-paratope} and structural \citep{antibodyipatch,daberdaku,pecan} approaches in prior art. The sequence-based models offer a strong and fast baseline in this setting, as the median gap between paratope residues in a CDR is 1--2 residues  \citep{motifpaper}---hence the paratope is reasonably sequentially located within the CDR.

On the other hand, there is no Fv-like region for epitopes, and as a result the \emph{entire} antigen needs to be considered. This gives epitope predictions a high degree of class imbalance: as no part of the antigen can be easily discarded at the input stage, most antigen residues will not belong to an epitope. Further, epitope residues may be far apart in the antigen sequence, implying that sequential epitope models are likely to be weak baselines. Epitopes are also specific to the antibody that targets them, implying that antibody information needs to be carefully leveraged within an epitope predictor \citep{epipred,discotope}.

\section{Present work}

Clear \emph{asymmetry} arises between paratope and epitope prediction: paratopes are highly sequential and can be predicted well in isolation, while epitopes are structural in nature and are inherently conditioned by the paratope. Hence, we conclude that a joint paratope-epitope predictor should ideally have separately tuned architectures for the two tasks. However, the present state-of-the-art for joint paratope-epitope prediction, PECAN \citep{pecan}, uses an entirely symmetric architecture for both.

Accordingly, here we propose \emph{epitope-paratope message passing} ({\bf EPMP}), which designs highly asymmetrical neural networks for paratope (\emph{Para-EPMP}) and epitope (\emph{Epi-EPMP}) predictors. 

Para-EPMP is designed to exploit the sequentiality of the paratope (using \`{a} trous convolutional neural networks), while also leveraging structural cues with graph neural networks (GNNs). We use the most expressive class of spatial GNN architectures to further take advantage of the rich and balanced paratope label information.

Conversely, Epi-EPMP is purely structural, and extensively enforces exploiting the contextual cues from the antibody, along with a multi-task approach that ensures the model's predictions are adequately regularised. Given the high potential for overfitting, we leverage the more versatile attentional class of GNNs.

\section{Data}
The preprocessed datasets are taken from PECAN  \citep{pecan}. Different datasets are used for paratope and epitope prediction, as complexes have been filtered to ensure that no two antibodies/antigens are similar in their respective datasets.

As in other works, the paratope is defined to include the residues on the antibody which are less than 4.5\r{A} away from any heavy atom in the antigen (non hydrogen atoms). The epitope is defined similarly, but for antigen residues.

The paratope dataset is a subset of the AbDb database \citep{abdb}, containing 308 complexes in the training and validation sets and 152 complexes in the test set. The Fv regions of the antibodies are used. In the test set, 9.4\% of antibody residues are part of the positive class.

The epitope dataset is a subset of the SAbDab database \citep{sabdab} and the Docking Benchmark v5 \citep{dbv5}. There are 132 complexes in the combined training/validation set, and 30 complexes in the test set. The CDR regions of the antibodies are used, to provide highly targeted context. In the test set, 7.8\% of antigen residues were part of the positive class.

The same features as in PECAN are used: a 62-dimensional feature vector for each residue, containing a one-hot encoding of the amino acid, alignment based features from the Position Specific Scoring Matrix (PSSM), surface accessibility scores and a profile counting the number of amino acid types in a sphere of radius 8\r{A} around the residue. The graphs in PECAN are restricted to a neighbourhood of the closest 15 neighbours (up to a distance of 10\r{A}). To allow these to be used as undirected graphs, we symmetrise the adjacency matrices. This diversifies the structure in the graph, as residue nodes will have varying numbers of neighbours.

\section{Methods}

We now present our two predictors within EPMP, by considering paratope and epitope prediction in turn.

\subsection{Paratope model (Para-EPMP)}
\begin{figure}[h!]
\vskip 0.2in
\begin{center}
\centerline{\includegraphics[width=\columnwidth]{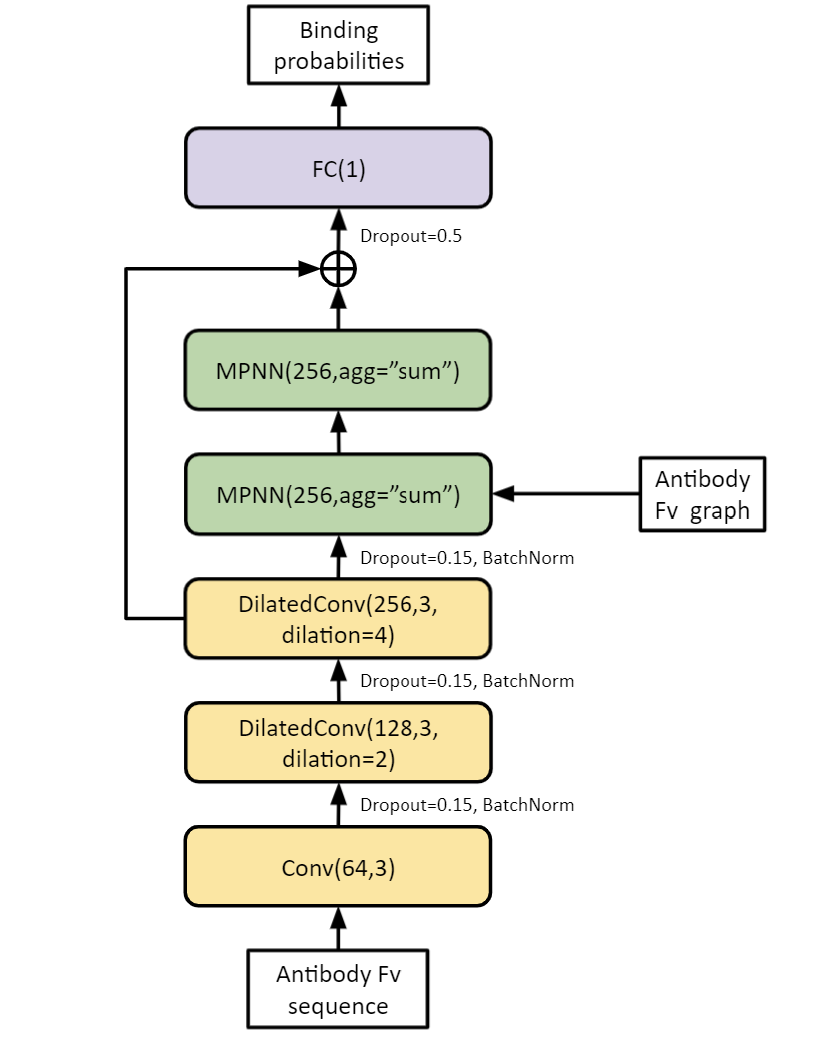}}
\caption{The Para-EPMP architecture. The output feature dimension for each layer is the first term in the bracket. For the convolutional layers, the second term is the kernel size.}
\label{fig:MultitaskP}
\end{center}
\vskip -0.2in
\end{figure}

For our paratope model, the input features are initially processed sequentially, with graph structure incorporated later. This model and its hyperparameters are outlined in Figure \ref{fig:MultitaskP}, and closely echoes our previous discussion.

The residue features are first passed through three \`{a} trous CNNs \citep{kalchbrenner2016neural}, as used in the sequence-based paratope predictor AG-Fast-Parapred \citep{agfastparapred}, as well as the recent Tail-GNN model \citep{spalevic2020hierachial}. \`{A} trous CNNs have a \emph{dilation} term which creates gaps in the convolution kernel, allowing the entire sequence to be covered in fewer layers, preventing overfitting.

The output embeddings from the CNN are then passed through two message passing neural network (MPNN) layers \citep{mpnn} over the proximity-based residue graph. These layers have single linear layers as their message and update functions and use the sum aggregation. MPNNs correspond to the most expressive class of spatial GNNs, making them highly appropriate for the kind of data available for the paratope task.

We further leverage a skip connection \citep{he2016deep} across the MPNNs, allowing the model to circumvent any graph-based processing. A final linear layer, with logistic sigmoid activations, produces the predictions.

\subsection{Epitope model (Epi-EPMP)}

Our epitope model, in contrast, is purely structural---uses only GNN layers---and incorporates substantial context from the antigen. Given the large potential for overfitting in this setting, we use two graph convolutional network (GCN) encoders from \citet{gcn} to process the antibody and antigen graphs in isolation. GCNs feature a fully nonparametric aggregation process, only learning point-wise shared linear transformations. This makes them resistant to overfitting, and scalable to larger graphs.

To combine information from the two representations, we allow the residue representations to interact across the antibody-antigen boundary. For this purpose, we used a graph attention network (GAT) \citep{gat} over the fully connected bipartite graph between the antigen residues and the antibody residues (i.e. each antigen residue is attending over every antibody residue, and vice-versa).

The attention coefficients are used to combine the neighbourhood of a node in a learned weighted manner. Within the GAT, we utilise edge dropout \citep{rong2019dropedge} to allow the network to learn a more robust neighbourhood to attend over. This is similar to the cross-modal attention mechanism of AG-Fast-Parapred \citep{agfastparapred}, but we do not feed any privileged information into its connections.

Once attention is performed, the network then predicts both the paratope and epitope residues at once. This paratope prediction is solely for aiding the epitope prediction and is not used in the final results. Without such a multi-task objective, antibody residues would remain unlabelled, and hence the GAT would have to learn in an unsupervised manner which antibody residues are the most important in the attention layers. Instead, by leveraging multitasking, the network is given explicit cues as to which parts of the antibody are actually relevant to the binding site. Furthermore, it allows for knowledge transfer between the two tasks, regularising the epitope representations. 

\begin{figure}[h!]
\vskip 0.2in
\begin{center}
\centerline{\includegraphics[width=\columnwidth]{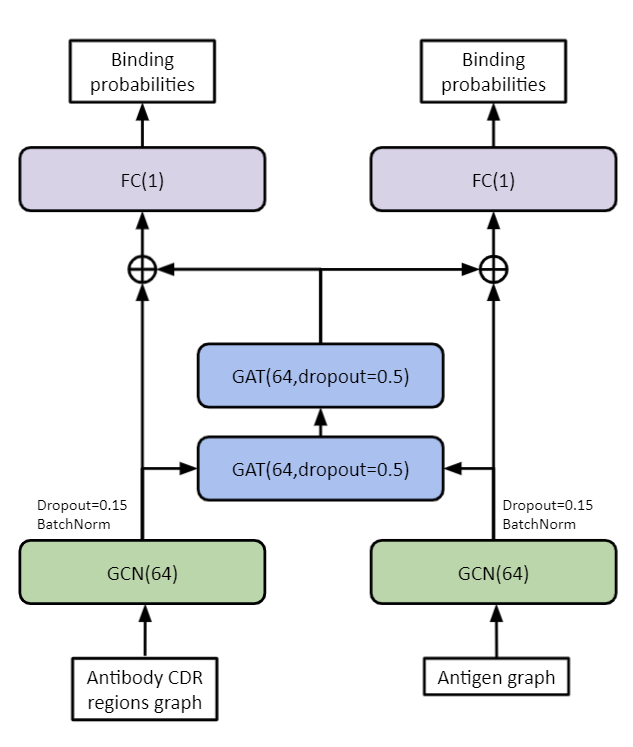}}
\caption{The Epi-EPMP multitask architecture. The output feature dimensions for each layer are given.}
\label{fig:MultitaskE}
\end{center}
\vskip -0.2in
\end{figure}

\subsection{Training}

The networks have been trained by optimising the class-weighted binary cross-entropy loss, which is common for imbalanced binary classification tasks. In the training phase for Epi-EPMP, the losses from both tasks (paratope and epitope prediction) are added together.

We optimise the cross-entropy using the Adam SGD optimiser \citep{kingma2014adam}, with an initial learning rate of $10^{-4}$ for Para-EPMP. To account for the unstable learning dynamics on Epi-EPMP, an initial learning rate of $10^{-5}$ was used to train its GCN layers, while an initial learning rate of $10^{-3}$ was used for all other layers.
\subsection{Evaluation}

Due to the large class imbalance, the area under the precision-recall curve (AUC PR) is the primary metric we use for benchmarking our models---matching the evaluation of PECAN. Prior art also reports the area under the receiver operating characteristic curve (AUC ROC), which we report as well for easier comparisons. The values reported are aggregated across five random seeds.

\section{Results}

\begin{table}[ht]
\caption{Models benchmarked on paratope prediction. The values from the first three models are reprinted from \citet{pecan}. Para-MPNN and Para-GCN are ablations of our model without the sequential (CNN) component, using either an MPNN or a GCN as the structural component.}
\label{paratope_bench}
\vskip 0.15in
\begin{center}
\begin{small}
\begin{sc}
\begin{tabular}{lcccr}
\toprule
Model & AUC ROC & AUC PR \\
\midrule

 Antibody i-Patch & 0.840&0.376\\
 Daberdaku et al. &0.950&0.658\\
 PECAN &0.957±0.000&0.700±0.001\\ \midrule
 Para-EPMP (ours)&\textbf{0.966±0.000}&\textbf{0.752±0.003}\\ \midrule
 Para-MPNN & 0.964±0.000 & 0.744±0.003\\
 Para-GCN & 0.956±0.000 & 0.681±0.004\\
\bottomrule
\end{tabular}
\end{sc}
\end{small}
\end{center}
\label{tbl:para}
\vskip -0.1in
\end{table}

\begin{table}[ht]
\caption{Models benchmarked on epitope prediction. The performance reported for PECAN is reprinted from \citet{pecan}. Epi-GCN and Epi-MPNN predict the epitope in a single-task setting (without the antibody), using either a GCN or an MPNN encoder on the antigen. Epi-CNN-GCN does the same, while incorporating a sequential (CNN) component in the same way as Para-EPMP.}
\label{Epitope_prediction}
\vskip 0.15in
\begin{center}
\begin{small}
\begin{sc}
\begin{tabular}{lcccr}
\toprule
Model & AUC ROC & AUC PR \\
\midrule
 PECAN & Not reported & 0.212±0.007\\ \midrule
 Epi-EPMP (ours) & \textbf{0.710± 0.003}& \textbf{0.277± 0.002}\\ \midrule
 Epi-GCN & 0.671±0.004 & 0.229±0.007\\
 Epi-MPNN &0.650±0.003 &0.206±0.005\\
 Epi-CNN-GCN &0.634±0.003 & 0.210±0.001\\
\bottomrule
\end{tabular}
\end{sc}
\end{small}
\end{center}
\label{tbl:epi}
\vskip -0.1in
\end{table}

The results from our quantitative evaluation are shown in Tables \ref{tbl:para}--\ref{tbl:epi}. While it may be observed that our model significantly outperforms the prior state-of-the-art (PECAN), it is important to emphasise the benefits of our \emph{asymmetric} approach. We additionally provide ablation studies of variants of our model that are less asymmetric, demonstrating that the exact combination of architectural changes we introduced is necessary to recover peak performance.

It is worth noting that PECAN also performs transfer learning experiments, using a dataset of general protein-protein interactions. While they report slightly improved results from this (epitope AUC PR: 0.242; paratope AUC PR: 0.703), our models are still outperforming, using only the original dataset. We defer investigations of transfer learning objectives for EPMP to future work.



\begin{figure}[h!]
\centering
\includegraphics[scale=0.42]{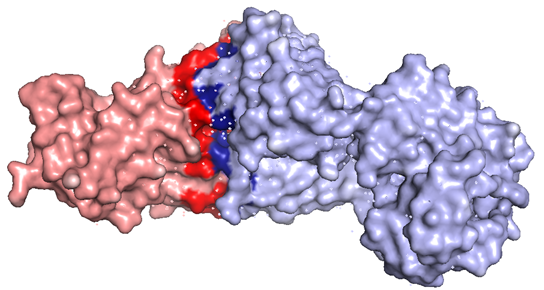}
\includegraphics[scale=0.42]{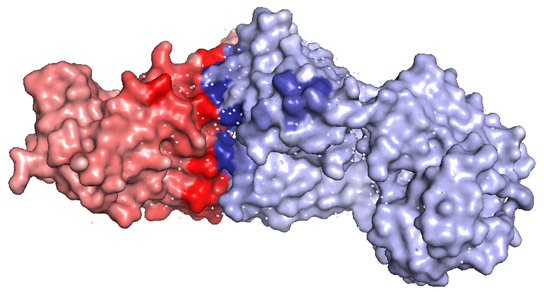}
\caption{B38-RBD (PDB: 7bz5). The red surface denotes the receptor binding domain of the SARS-CoV-2 spike protein, and the blue surface denotes its corresponding antibody (B38). {\bf Top}: Ground-truth binding interface. Darker residues denote the binding region. {\bf Bottom}: Predicted binding interface from running Para-EPMP on the antibody and Epi-EPMP on the antigen. Probabilities are coloured on a gradient scale, where the darkest residues are given the highest probability of participating in the binding.}
\label{fig:truebindingcovid}
\end{figure}

\paragraph{Qualitative study}
While there is substantial ground left to cover for robust epitope predictions, we demonstrate that our joint predictor can already produce actionable predictions, especially when the antigen is small. We evaluated EPMP's ability to predict the binding interface between a COVID-19 neutralising antibody (B38) and the RBD (Receptor Binding Domain) of the Spike protein, S1 of SARS-CoV-2 \citep{covid7bz5}. From Figure \ref{fig:truebindingcovid}, we can observe that the model is able to identify the correct localised region for the epitope, with any false positives being close to the binding interface.

\newpage
\bibliographystyle{icml2021}
\bibliography{example_paper}

\end{document}